# SPHEROIDAL QUANTUM WELL

N.A. Usov[a)]

*Pushkov Institute of Terrestrial Magnetism, Ionosphere and Radio Wave Propagation, Russian Academy of Sciences, IZMIRAN, 108480, Troitsk, Moscow, Russia*

Standard power series are used to construct and analyze angular and radial spheroidal functions, which are necessary for solving boundary value problems for Helmholtz equation in a spheroid. With an advanced approach the low-lying energy levels of a deep spheroidal quantum well are calculated as a function of the spheroid semiaxes ratio $a/b$. The well-known results for cylindrical and spherical wells are reproduced in the limits $a/b \gg 1$, and $a/b \to 1$, respectively.

## I. INTRODUCTION

The study of quantum states in fine metallic and semiconductor particles, as well as in semiconductor quantum dots, is important for the development of many branches of modern condensed matter physics. In addition, the construction of the energy spectrum of a quantum particle in a potential well is one of the first problems faced by a student starting to study quantum mechanics. The solution of the Schrödinger equation for cylindrical and spherical quantum wells in the corresponding orthogonal coordinate systems is given in most standard textbooks[1-5] on quantum mechanics. Meanwhile, the theory of spheroidal functions, which are necessary for solving the Dirichlet and Neumann boundary value problems in a spheroid, is presented in numerous monographs[6-9], reference books[10-11] and original articles[12-19] in a rather cumbersome mathematical language. The study of these works requires great effort and time. In this paper the standard power series are used to construct and study angular and radial spheroidal functions, which allows solving some boundary value problems for the Helmholtz equation in a spheroid at a level acceptable in university teaching. Using this approach the low-lying energy levels of a deep spheroidal quantum well are calculated as a function of the spheroid semiaxes ratio.

## II. ANGULAR AND RADIAL SPHEROIDAL FUNCTIONS

Let $a > b$ be the semiaxes of a prolate spheroid stretched along the Z axis, whose surface in the Cartesian coordinates is given by the equation

$$\frac{x^2 + y^2}{b^2} + \frac{z^2}{a^2} = 1. \tag{1}$$

For a lot of problems in physics and technology it is necessary to solve the Helmholtz equation in the volume of a spheroid

$$\Delta u + q^2 u = 0, \tag{2}$$

with simple boundary conditions on the spheroid surface $S$

$$u|_S = 0; \qquad \text{or} \qquad \frac{\partial u}{\partial n}\bigg|_S = 0, \tag{3}$$

where $n$ is the direction of the normal to the spheroid surface.

Equation (2) allows separation of variables in prolate spheroidal coordinates $(\xi, \eta, \varphi)$[6-9]

$$x = \rho \cos\varphi; \quad y = \rho \sin\varphi; \quad z = \frac{d}{2}\xi\eta; \quad \rho = \frac{d}{2}\sqrt{(\xi^2-1)(1-\eta^2)}, \tag{4}$$

where $\xi \in [1,\infty)$, $\eta \in [-1,1]$, $\varphi \in [0,2\pi]$, and $d = 2\sqrt{a^2 - b^2}$ is the distance between the foci of the spheroid section in the *XZ* plane. In these variables the function $u$ satisfies the equation

$$\frac{\partial}{\partial \xi}\left((\xi^2-1)\frac{\partial u}{\partial \xi}\right) + \frac{\partial}{\partial \eta}\left((1-\eta^2)\frac{\partial u}{\partial \eta}\right) + \frac{\xi^2-\eta^2}{(\xi^2-1)(1-\eta^2)}\frac{\partial^2 u}{\partial \varphi^2} + c^2(\xi^2-\eta^2)u = 0. \tag{5}$$

where the dimensionless parameter $c = qd/2$. The solution to Eq. (5) can be sought as

$$u(\xi,\eta,\varphi) = R(\xi)S(\eta)e^{im\varphi}, \tag{6}$$

where $m = 0, \pm 1, \pm 2, \ldots$ since the function (6) has to be single valued with respect to the variable $\varphi$.

Substituting (6) into Eq. (5), one obtains the system of equations to determine the angular $S(\eta)$ and radial $R(\xi)$ spheroidal functions

$$\frac{d}{d\eta}\left((1-\eta^2)\frac{\partial S}{\partial \eta}\right) + \left(\lambda - c^2\eta^2 - \frac{m^2}{1-\eta^2}\right)S = 0; \tag{7a}$$

$$\frac{d}{d\xi}\left((\xi^2-1)\frac{\partial R}{\partial \xi}\right) - \left(\lambda - c^2\xi^2 + \frac{m^2}{\xi^2-1}\right)R = 0, \tag{7b}$$

where $\lambda$ is the separation constant.

An interesting feature of Eqs. (7) is the fact that the separation constant $\lambda$ is actually a function of the dimensionless constant $c$ related to the eigenvalue $q$ of Eq. (2). The dependence $\lambda = \lambda(c)$ is determined from the condition of regular behavior of the function $S(c,\eta)$ near its singular points, $\eta = \pm 1$, which correspond to the spheroid poles. To solve the internal boundary value problem in a spheroid, the regular behavior of the function $R(c,\xi)$ at the point $\xi = 1$ is also necessary.

It follows from the definition of spheroidal coordinates (4) that the surface of a spheroid is given by the equation $\xi = a/\sqrt{a^2 - b^2} = 2a/d$. Therefore, boundary conditions (3) take the form

$$R(c,\xi)\big|_{\xi=2a/d} = 0, \quad \text{or} \quad \frac{dR}{d\xi}(c,\xi)\bigg|_{\xi=2a/d} = 0. \tag{8}$$

The eigenvalues of the corresponding boundary value problem are found from the relation $q = 2c/d$.

Let us begin the analysis of equation (7a) with the simplest case $m = 0$. By definition, the angular function $S(c,\eta)$ is given in the interval $|\eta| \leq 1$. In the standard approach[6-11], Eq. (7a) is solved by expanding $S(c,\eta)$ into a series of Legendre polynomials. However, it leads to cumbersome recurrent relations. In fact, in the limited area $|\eta| \leq 1$ it is reasonable to look for a power series solution to Eq. (7a)

$$S(c,\eta) = a_0 + a_1\eta + a_2\eta^2 + a_3\eta^3 + \ldots \tag{9}$$

The coefficients of series (9) for even and odd solutions to equation (7a) satisfy the recurrent relations (10a) and (10b), respectively

$$a_0 = 1; \quad a_2 = -\lambda a_0/2; \quad a_{2k+4} = \frac{(2k+2)(2k+3) - \lambda}{(2k+3)(2k+4)}a_{2k+2} + \frac{c^2}{(2k+3)(2k+4)}a_{2k}, \tag{10a}$$

$$a_1 = 1; \quad a_3 = (2-\lambda)a_1/6; \quad a_{2k+5} = \frac{(2k+3)(2k+4) - \lambda}{(2k+4)(2k+5)}a_{2k+3} + \frac{c^2}{(2k+4)(2k+5)}a_{2k+1}, \tag{10b}$$

where $k = 0, 1, 2, \ldots$. The coefficients $a_0$ and $a_1$ which remain undefined without loss of generality can be set equal to one. Thus, the series (9), (10) are completely defined. It is easy to see that in the limit $k \to \infty$ the successive coefficients of the series (9), (10) satisfy the relations

$$a_{2k+2}/a_{2k} \to 1, \quad a_{2k+3}/a_{2k+1} \to 1. \tag{11}$$

Therefore, power series (9), (10) certainly converge in the open interval $-1 < \eta < 1$ for any $\lambda$ and $c$ values. However, they generally diverge at singular points $\eta = \pm 1$. The true value of the separation constant $\lambda = \lambda(c)$ is determined by the condition that the series (9), (10) converge at the spheroid poles as well.

Similar solution for the function $S(c,\eta)$ can be constructed also for the case $m \geq 1$. Making in Eq. (7a) the substitution[6,9]

$$S = (1-\eta^2)^{m/2}v(\eta), \tag{12a}$$

one obtains for the function $v(\eta)$ the equation

$$(1-\eta^2)\frac{d^2v}{d\eta^2} - 2(m+1)\eta\frac{dv}{d\eta} + \left(\lambda - m(m+1) - c^2\eta^2\right)v = 0. \tag{12b}$$

The latter is also solved by expanding $v(\eta)$ in a power series. The coefficients of even and odd solutions of Eq. (12b) are given by formulas (13a) and (13b), respectively.

$$a_0 = 1; \qquad a_2 = \frac{m(m+1)-\lambda}{2}a_0;$$

$$a_{2k+4} = \frac{(2k+2)(2k+3+2m)+m(m+1)-\lambda}{(2k+3)(2k+4)}a_{2k+2} + \frac{c^2}{(2k+3)(2k+4)}a_{2k}; \tag{13a}$$

$$a_1 = 1; \qquad a_3 = \frac{(m+1)(m+2)-\lambda}{6}a_1;$$

$$a_{2k+5} = \frac{(2k+3)(2k+4+2m)+m(m+1)-\lambda}{(2k+4)(2k+5)}a_{2k+3} + \frac{c^2}{(2k+4)(2k+5)}a_{2k+1}, \tag{13b}$$

where $k = 0, 1, 2, \ldots$ Obviously, in the case $m \geq 1$ the successive coefficients of series (13) also satisfy relations (11) in the limit $k \to \infty$. Note, that $m$ must be replaced by $|m|$ in Eqs. (12), (13) for negative integers $m$. The orthogonality of eigenfunctions (6) with quantum numbers $m$ and $-m$ is guaranteed by the phase factor $\exp(im\varphi)$.

Power expansions (9)-(13) are well known and are given in manuals[9,11], in particular. In our opinion, formulas (9) - (13) are the most convenient for constructing and studying angular spheroidal functions. Moreover, they can also be used to determine the separation constant $\lambda$ as a function of the parameter $c$.

Let us consider first the case $c = 0$, when the Helmholtz equation (2) turns into the Laplace equation. As is known[6], in this case, recurrence relations (13) can be terminated at $\lambda_k^{ev} = (2k+2)(2k+3+2m)+m(m+1)$ for even functions, and at $\lambda_k^{odd} = (2k+3)(2k+4+2m)+m(m+1)$ for odd functions, respectively, where $k = -1, 0, 1, 2, \ldots$ As a result, the angular eigenfunctions of Laplace equation are proportional to Legendre polynomials, or associated Legendre functions for $m = 0$ and $m \geq 1$, respectively. These functions have finite values at the spheroid poles, at $\eta = \pm 1$. For other values of $\lambda$ the power series (13) do not terminate, and due to the limiting relations (11), they diverge at the spheroid poles.

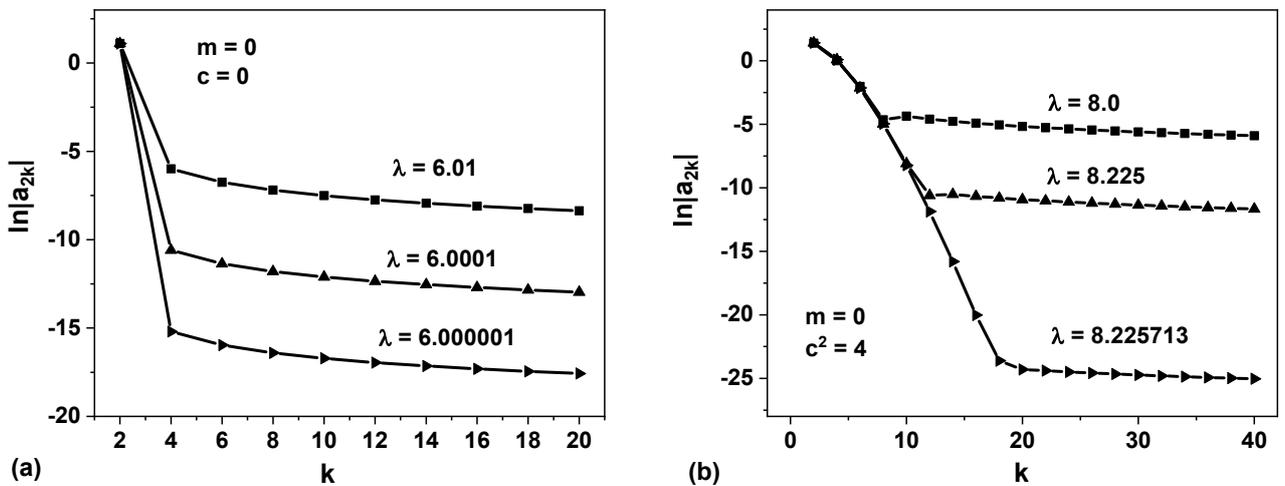

Fig. 1. a) Behavior of the coefficients $a_{2k}$ of Eq. (10a) for the angular function $S(0,\eta)$ in the case of $m = 0$, $c = 0$ near the exact value of the separation constant $\lambda = 6$. b) The same for the case of $m = 0$, $c^2 = 4$ near the tabular value[11] $\lambda = 8.225713$.

It is useful to analyze the behavior of the coefficients of power series (13) for the exactly solvable case $c = 0$ in a situation where the separation constant $\lambda$ changes near some exact eigenvalue. As a simple example, Fig. 1a shows the behavior of the coefficients $a_{2k}$ for the case $m = 0$ as $\lambda$ changes near one of the exact eigenvalues, $\lambda = 6$. In the limit $\lambda \to 6$ the function $S(0,\eta)$ given by formulas (9), (10a) approaches the exact eigenfunction $1 - 3\eta^2$, but at the same time it retains an infinitely long tail of very small, but due to Eq. (11), slowly decreasing terms. This leads to the divergence of the series at the spheroid poles.

Note that the coefficients of the series (10a) demonstrate the same behavior on large numbers $k$ in the general case, $c > 0$. Consider for example the case $m = 0$, $c^2 = 4$. One finds in Table 21.1 of Ref. 11 that in this case the eigenvalue for the second eigenfunction is given by $\lambda = 8.225713$. As Fig. 1b shows, when $\lambda$ approaches to the tabular value, the coefficients $a_{2k}$ very quickly reach the same asymptotic behavior with increasing $k$. This asymptotic behavior actually coincides with the series given by the relation

$$a_{2k+4} = \frac{k+1}{k+2} a_{2k+2}. \tag{14}$$

Based on this observation, one can conclude that the condition $a_k(\lambda) = 0$ can be considered as an approximate equation for the separation constant $\lambda$. Of course, the larger the number of the coefficient $k$, the more accurate values of $\lambda$ will be determined from this equation.

It follows from Eqs. (10) or (13) that for a given value of $c$ each coefficient $a_k(\lambda)$ is, in fact, some polynomial in $\lambda$. It can be shown that the roots of these polynomials are simple. Let the polynomial $a_{2k}(\lambda)$ given by Eq. (13a) has some root $\lambda_{2k}$, and the next polynomial $a_{2k+2}(\lambda)$ has a close root $\lambda_{2k+2}$. Then, in some neighborhood of these roots, both polynomials can be represented as

$$a_{2k}(\lambda) = A_{2k}(\lambda - \lambda_{2k}); \qquad a_{2k+2}(\lambda) = A_{2k+2}(\lambda - \lambda_{2k+2}),$$

where the constants $A_{2k}$ and $A_{2k+2}$ are the derivatives of these polynomials calculated at the given roots.

Consider now the equation for the following coefficient $a_{2k+4}(\lambda)$

$$a_{2k+4} = \frac{(2k+2)(2k+3+2m) + m(m+1) - \lambda}{(2k+3)(2k+4)} A_{2k+2}(\lambda - \lambda_{2k+2}) + \frac{c^2}{(2k+3)(2k+4)} A_{2k}(\lambda - \lambda_{2k}). \tag{15}$$

Let us show that the polynomial (15) has a root close to the roots $\lambda_{2k}$, $\lambda_{2k+2}$. Equating relation (15) to zero, we obtain the quadratic equation

$$\lambda^2 - [(2k+2)(2k+3+2m) + m(m+1) + \lambda_{2k+2} + \gamma]\lambda + [(2k+2)(2k+3+2m) + m(m+1)]\lambda_{2k+2} + \gamma\lambda_{2k} = 0,$$

where we put $\gamma = c^2 A_{2k}/A_{2k+2}$. Solving this equation under the condition

$$(2k+2)(2k+3+2m) + m(m+1) \gg \lambda_{2k+2}, \gamma, \gamma\lambda_{2k},$$

one obtains two roots

$$\lambda_a = (2k+2)(2k+3+2m) + m(m+1) + \gamma + \frac{\lambda_{2k+2} - \lambda_{2k}}{(2k+2)(2k+3+2m) + m(m+1)}\gamma + O\left(\frac{1}{(2k+2)^2(2k+3)^2}\right); \tag{16}$$

$$\lambda_b = \lambda_{2k+2} - \frac{\lambda_{2k+2} - \lambda_{2k}}{(2k+2)(2k+3+2m) + m(m+1)}\gamma + O\left(\frac{1}{(2k+2)^2(2k+3)^2}\right). \tag{17}$$

The root $\lambda_a$, Eq.(16), is quite large for $(2k)^2 \gg 1$ and is out of interest. However, it follows from Eq. (17) that in the limit $(2k)^2 \gg 1$ the root $\lambda_{2k+4} = \lambda_b$ is very close to the root $\lambda_{2k+2}$. Since the isolated roots of the successive polynomials $a_{2k+4}(\lambda)$ and $a_{2k+2}(\lambda)$ are very close each other at $(2k)^2 \gg 1$, the approximate values of the separation constant $\lambda$, at least for the eigenfunctions with moderate numbers, can be obtained by equating to zero any of the polynomials $a_{2k}(\lambda)$ with a sufficiently large index $2k$.

Let us assume that for some eigenfunction $S(c,\eta)$ the root $\lambda^* = \lambda_{2k+2}(c)$ has been taken as an approximate value for the separation constant. Then the coefficient $a_{2k+2}(\lambda^*)$ vanishes exactly. Furthermore, it follows from Eq. (15) that the coefficient $a_{2k+4}(\lambda^*)$ and all subsequent coefficients do not exceed a small value $\mu/(2k)^2$, where $\mu$ is some constant. Therefore, by choosing the index $2k$ large enough, the remainder of the series for this eigenfunction can be made arbitrarily small

$$|\Delta S(c,\eta)| = \left| \sum_{s=2}^{\infty} a_{2k+2s}(\lambda^*)\eta^{2k+2s} \right| \leq \frac{\mu}{(2k)^2} \eta^{2k} \sum_{s=2}^{\infty} \eta^{2s} < \frac{\mu}{(2k)^2} \frac{\eta^{2k}}{1-\eta^2} \qquad (18)$$

outside any finite neighborhood of points $\eta = \pm 1$. At the points $\eta = \pm 1$ themselves, the estimate (18) diverges, but this is not of great importance from a practical point of view.

The roots of the polynomials $a_k(\lambda)$, that is, the approximate values $\lambda^*(c)$, can be found by simple segment bisection method[20]. For a given $m$, the successive roots of the equation $a_k(\lambda) = 0$ will be denoted as $\lambda_{ml}$, where the index $l$ runs through the values $l = m, m+1, \ldots$ As initial values for the segment bisection method one can rely on the known values of the separation constant at $c = 0$, since it is well known[8,9,11] that the eigenvalues $\lambda(c)$ are monotonic functions of the parameter $c$, and the eigenvalues of successive eigenfunctions are well separated.

In Tables 1a-1c we compare the separation constants calculated for several angle eigenfunctions by the segment bisection method with the corresponding tabular values given in the handbook[11]. For even angular functions the values of the separation constant $\lambda_{ml}$ are obtained by determining the lowest positive roots of the polynomial $a_k(\lambda)$ at $k = 100$ at the 40-th iteration of the segment bisection method. For the odd angular functions, these values are obtained by analyzing the roots of the polynomial $a_{101}(\lambda)$. In all cases, the $\lambda_{ml}$ values are obtained with a ten-digit precision, and the time for determining an individual root takes a fraction of a second on a PC.

**Table 1a.** ($\lambda_{ml}$ at $c^2 = 1$, $m = 0$)

|  | $l = 0$ | $l = 1$ | $l = 2$ | $l = 3$ | $l = 4$ |
|---|---|---|---|---|---|
| Ref. 11 | 0.319000 | 2.593084 | 6.533471 | 12.514462 | 20.508274 |
| $a_k(\lambda) = 0$ | 0.31900005515 | 2.59308457998 | 6.53347180052 | 12.51446214509 | 20.50827436257 |

**Table 1b.** ($\lambda_{ml}$ at $c^2 = 4$, $m = 0$)

|  | $l = 0$ | $l = 1$ | $l = 2$ | $l = 3$ | $l = 4$ |
|---|---|---|---|---|---|
| Ref. 11 | 1.127734 | 4.287128 | 8.225713 | 14.100203 | 22.054829 |
| $a_k(\lambda) = 0$ | 1.12773406485 | 4.28712854396 | 8.22571300111 | 14.10020387620 | 22.05482977047 |

**Table 1c.** ($\lambda_{ml}$ at $c^2 = 2$, $m = 1$)

|  | $l = 1$ | $l = 2$ | $l = 3$ | $l = 4$ | $l = 5$ |
|---|---|---|---|---|---|
| Ref. 11 | 2.382655 | 6.841718 | 12.937881 | 20.965685 | 30.977891 |
| $a_k(\lambda) = 0$ | 2.38265540616 | 6.84171800891 | 12.93788107437 | 20.96568591695 | 30.97789126965 |

Fig. 2 shows the angular spheroidal functions at $m = 0$, $c^2 = 1$ and at $m = 1$, $c^2 = 2$, respectively. The corresponding separation constants $\lambda_{ml}$ are taken from Tables 1a-1c. Note that the number of zeros of the angular eigenfunctions within the interval $-1 < \eta < 1$ is given by $l - m$.

Let us now consider the solution of Eq. (7b) for the radial spheroidal function, taking into account that the separation constant $\lambda(c)$ is the same for Eqs. (7a) and (7b). It is easy to see that Eqs. (7a) and (7b) for the angular and radial functions actually coincide. Therefore, the same recurrent relations (10), (13) are also valid for the radial equation. However, in the region $\xi \geq 1$, where the radial Eq. (7b) is given, series (10), (13) diverge.

For arbitrary integers $m \geq 0$ it is useful to make in Eq. (7b) the substitution

$$R(\xi) = (\xi^2 - 1)^{m/2} v(\xi).$$

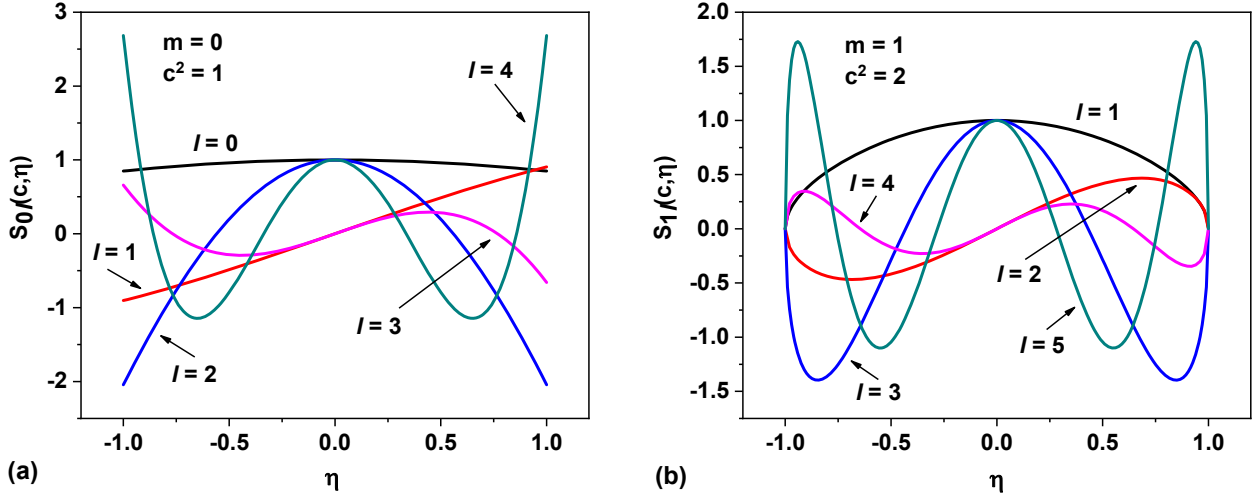

Fig. 2. Angular spheroidal functions for $m = 0$, $c^2 = 1$ (a) and $m = 1$, $c^2 = 2$ (b) depending on the eigenfunction number $l$.

The function $v(\xi)$ satisfies the same equation (12b), but for $\xi \geq 1$. To expand the function $v(\xi)$ in a power series, it is reasonable to introduce a new variable by setting $\xi = 1 + t$, $t \geq 0$. This leads to the equation

$$t(t+2)\frac{d^2v}{dt^2} + 2(m+1)(t+1)\frac{dv}{dt} + \left(c^2(t+1)^2 + m(m+1) - \lambda\right)v(t) = 0. \tag{19}$$

One can seek for a power series solution to this equation

$$v(c,t) = a_0 + a_1 t + a_2 t^2 + a_3 t^3 + \ldots \tag{20}$$

The coefficients of this series satisfy the recurrent relations

$$a_1 = \frac{\lambda - m(m+1) - c^2}{2(m+1)} a_0; \quad a_2 = \frac{1}{4(m+2)}\left[(\lambda - c^2 - (m+1)(m+2))a_1 - 2c^2 a_0\right], \tag{21a}$$

and in general

$$a_{k+3} = -\frac{1}{2(k+3)(k+3+m)}\left[((k+2)(k+3+2m) + m(m+1) + c^2 - \lambda)a_{k+2} + 2c^2 a_{k+1} + c^2 a_k\right], \tag{21b}$$

where $k = 0, 1, 2, \ldots$ The coefficient $a_0$ remains undefined, and without loss of generality can be set equal to one.

Note that for large $k$, the ratio of successive coefficients of series (20) tends to ½. Therefore, the power series (20), (21) is convergent at least in the interval $0 \leq t < 2$, that is, for $1 \leq \xi < 3$. This interval is sufficient for solving internal boundary value problems for the Helmholtz equation in spheroids with the semiaxes ratio $a/b > 3/2\sqrt{2} \approx 1.06$. Indeed, for many applications, it is sufficient to construct the regular function $R(c,\xi)$ at least in the interval $1 \leq \xi < 3$, since here lie the values of the variable $\xi^* = a/\sqrt{a^2 - b^2}$ that correspond to the positions of the surface of the prolate spheroids indicated.

Fig. 3 shows the radial functions $R(c,\xi)$ for $m = 0$ and 1 for various values of the parameter $c^2 = 1 - 4$. The behavior of these functions is regular in the interval $1 \leq \xi < 3$, where the series (20), (21) certainly converges.

Having determined the separation constants $\lambda_{ml}$ using the procedure described above, and constructed radial functions by means of Eqs. (20), (21) one can solve internal boundary value problems for the Helmholtz equation (2), (3) in a spheroid. As an example, in the next section the low-lying energy levels of a quantum particle in a spheroidal potential well of a sufficiently large depth are determined.

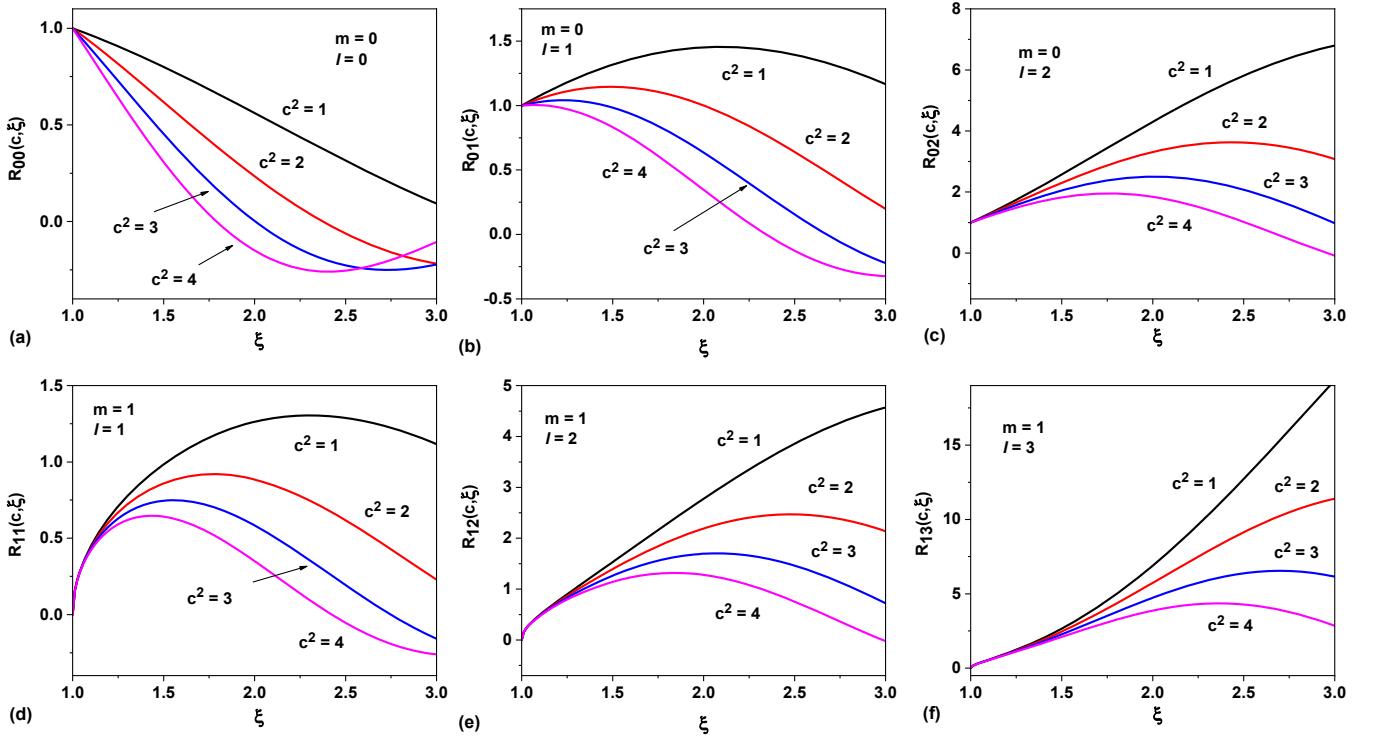

Fig. 3. Radial functions $R_{0l}(c,\xi)$ and $R_{1l}(c,\xi)$ in the interval $1 \le \xi < 3$ for different values of parameter $c^2 = 1 - 4$.

## III. SPHEROIDAL QUANTUM WELL

The Schrödinger equation and the boundary condition for the wave function of a quantum particle of mass $m$ in a spheroidal quantum well of a large depth are given by[3,4]

$$-\frac{\hbar^2}{2m}\Delta\Psi = E\Psi, \qquad \Psi|_S = 0. \qquad (22a)$$

Writing the particle energy as

$$E = \frac{\hbar^2}{2mb^2}(qb)^2, \qquad (22b)$$

where $b$ is the minor spheroid semiaxis, one arrives to Helmholtz equation (2), (3) to determine the eigenvalues $q$ of the boundary value problem (22). This can be done using the approach developed in Sec. II.

For a spheroidal quantum well the energy levels depend on the absolute value $|m|$ of the azimuthal quantum number, although the states with quantum numbers $m$ and $-m$ remain degenerate. The quantum number, which numerates the eigenfunctions at a given $m$, is still denoted by the index $l = |m|, |m|+1, \ldots$. The corresponding angular eigenfunctions are denoted by $S_{ml}(c,\eta)$, $l - |m|$ being the number of zeros of the angular function inside the segment (-1,1). Consecutive zeros of the radial function $R_{ml}(c,\xi)$ in the interval $\xi > 1$ are denoted as $\xi_i$, $i = 1, 2, 3, \ldots$ and numbered with the index $n_r = 1, 2, 3, \ldots$

The reduced eigenvalues $qb$ of boundary value problem (22) are calculated as follows. First, by analyzing the angular Eq. (7a), the dependence of the separation constants $\lambda_{ml}$ on the parameter $c$ is found for several values $m \ge 0$. Having determined the functions $\lambda_{ml}(c)$, one can use Eqs. (20), (21) to construct the radial functions $R_{ml}(\xi)$ on the segment $1 \le \xi < 3$, and to determine the positions of their consecutive zeros $\xi_1, \xi_2, \xi_3$, etc. The equality $R_{ml}(\xi_i) = 0$ means that the first boundary condition (8) is satisfied. Every root $\xi_i(m,l,c)$ of the radial function fixes the position of the spheroid boundary and simultaneously sets its aspect ratio, since

$$\frac{a}{b} = \frac{\xi_i}{\sqrt{\xi_i^2 - 1}}. \qquad (23)$$

On the other hand, since the parameter $c$ is related to the eigenvalue $q$ by the relation $c = q\sqrt{a^2 - b^2}$, one can calculate the reduced eigenvalues $qb$ for given $m$ and $l$ as

$$qb = \frac{c}{\sqrt{a^2/b^2 - 1}} = c\sqrt{\xi_i^2 - 1}. \quad (24)$$

The results of the calculations of the separation constants and low-lying eigenvalues $qb$ of spheroidal quantum well depending on the spheroid aspect ratio $a/b$ are shown in Fig. 4.

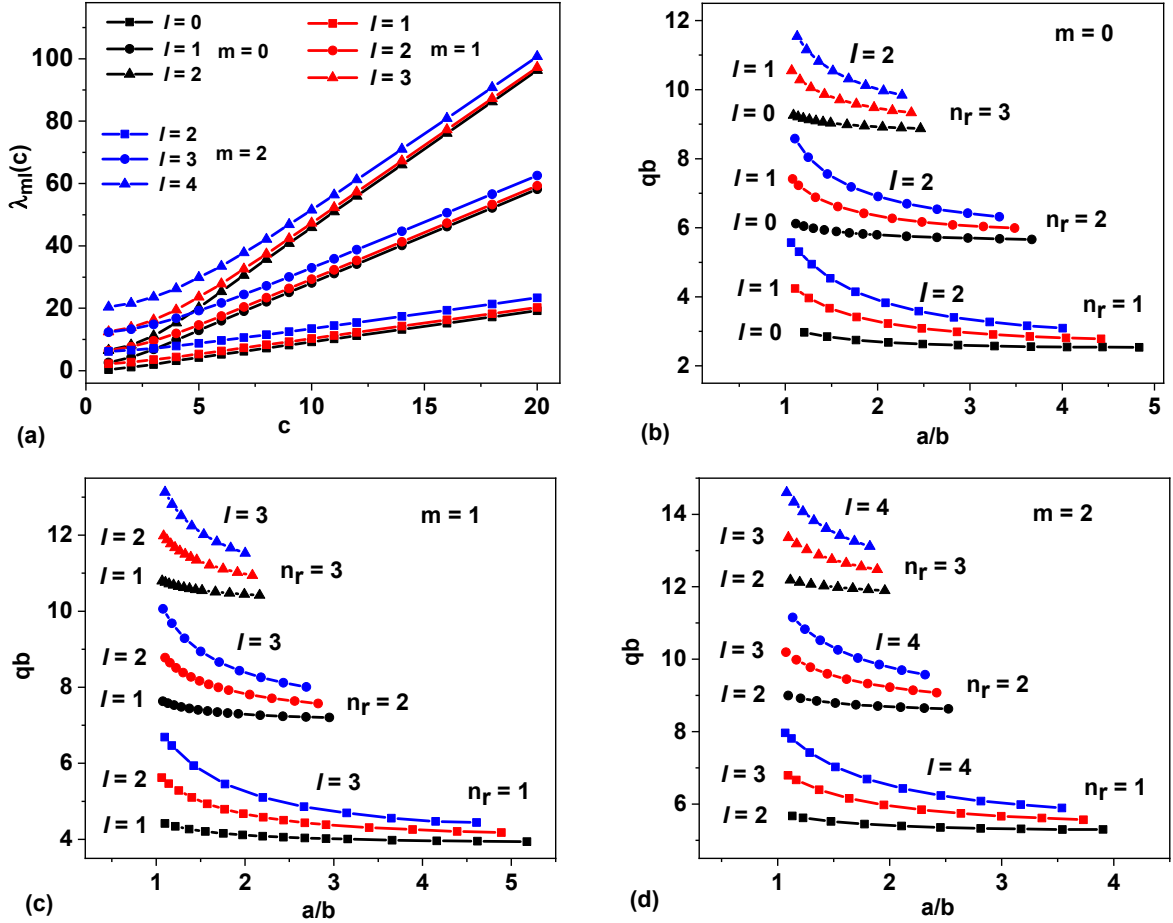

Fig. 4. (a) Separation constants $\lambda_{ml}$ versus parameter $c$ for low-lying quantum states of a deep spheroidal potential well. (b) - (d) Reduced eigenvalues $qb$ defined in Eq. (22b) as the functions of the spheroid aspect ratio $a/b$ for the cases $m = 0$, $m = 1$, and $m = 2$, respectively.

It is interesting to compare the energy spectrum of a spheroid, presented in Fig. 4, with the well-known limiting cases of an infinite cylinder ($a/b \to \infty$) and a sphere ($a/b = 1$). Obviously, in the limit $a/b \gg 1$ one obtains a cylindrical potential well, for which the solutions of the quantum problem (22) are the wave functions

$$\Psi(\rho, \varphi, z) = \Phi(\rho)\exp(im\varphi)\exp(ipz), \quad (25)$$

the radial function $\Phi(\rho)$ being the solution of the Bessel equation regular at zero

$$\frac{1}{\rho}\frac{d}{d\rho}\left(\rho\frac{d\Phi(\rho)}{d\rho}\right) + \left(q^2 - p^2 - \frac{m^2}{\rho^2}\right)\Phi(\rho) = 0, \quad (26)$$

i.e. $\Phi(\rho) = J_m\left(\sqrt{q^2 - p^2}\,\rho\right)$. For low-lying quantum states in a cylinder the quantum number $p \to 0$. In this limit the reduced eigenvalues $qR$, where $R$ is the radius of the cylinder, are the roots of the equations $J_m(x) = 0$[3,4]. The successive roots of the radial equation (26) are also numbered by the integers $n_r = 1, 2, 3$, etc.

As Figs. 4b-4d show, in the limit $a/b \gg 1$ for various values $n_r = 1, 2, 3$ the reduced eigenvalues of the spheroidal quantum well with different $l$ cluster near the levels with $l - m = 0$. The corresponding angular spheroidal functions have no zeros inside the segment $(-1,1)$. As Table 2 shows, in the limit $a/b \gg 1$, the eigenvalues $qb$ of a spheroidal well with quantum numbers $l - m = 0$ approach the eigenvalues $qR$ of a cylindrical well with the same quantum numbers $m$ and $n_r$. Some difference in the numerical values for $qb$ and $qR$ in Table 2 is associated with the finiteness of the aspect ratio of the spheroid, which does not exceed 5 in the calculations performed.

**Table 2**.

|  | $m = l = 0$ | $J_0(x) = 0$ | $m = l = 1$ | $J_1(x) = 0$ | $m = l = 2$ | $J_2(x) = 0$ |
|---|---|---|---|---|---|---|
|  | $qb, a/b \gg 1$ | $qR$ | $qb, a/b \gg 1$ | $qR$ | $qb, a/b \gg 1$ | $qR$ |
| $n_r = 1$ | 2.53676 | 2.4048 | 3.93795 | 3.8317 | 5.29869 | 5.1356 |
| $n_r = 2$ | 5.66196 | 5.5201 | 7.20434 | 7.0156 | 8.62876 | 8.4172 |
| $n_r = 3$ | 8.87324 | 8.6537 | 10.42078 | 10.1734 | 11.89508 | 11.6198 |

Let us now compare the energy levels in a spheroid with $a/b \approx 1$ and a spherical particle with $a/b = 1$. This case is especially interesting, since the energy levels in a sphere are degenerate[1-4] in the quantum number $m$. Therefore, the quantum numbers $n_r$ and $l$ are used to enumerate the energy levels in a sphere. At the same time, as Fig. 4 shows, the energy levels in a spheroid depend on the quantum number $m$ appreciably. From topological considerations, it is clear that for a sphere and a spheroid with $a/b \approx 1$, the number of nodal lines of eigenfunctions in radial and angular variables must match. Therefore, similar wave functions should have the same number of zeros $n_r$ for radial functions and the same number of internal zeros $l - |m|$ for angular functions.

**Table 3**

|  |  | $n_r = 1$, sphere | | $n_r = 1$, spheroid, $a/b \approx 1$ | |
|---|---|---|---|---|---|
| $l = 0$ | $m = 0$ | $qR = 3.142$ | $n_\theta = 0$ | $qb = 2.97214$ | $n_\eta = 0$ |
| $l = 1$ | $m = 0$ | $qR = 4.493$ | $n_\theta = 1$ | $qb = 4.23953$ | $n_\eta = 1$ |
|  | $m = \pm 1$ | $qR = 4.493$ | $n_\theta = 0$ | $qb = 4.41557$ | $n_\eta = 0$ |
| $l = 2$ | $m = 0$ | $qR = 5.764$ | $n_\theta = 2$ | $qb = 5.56938$ | $n_\eta = 2$ |
|  | $m = \pm 1$ | $qR = 5.764$ | $n_\theta = 1$ | $qb = 5.62226$ | $n_\eta = 1$ |
|  | $m = \pm 2$ | $qR = 5.764$ | $n_\theta = 0$ | $qb = 5.6712$ | $n_\eta = 0$ |
| $l = 3$ | $m = 0$ | $qR = 6.988$ | $n_\theta = 3$ | $qb = 6.63587$ | $n_\eta = 3$ |
|  | $m = \pm 1$ | $qR = 6.988$ | $n_\theta = 2$ | $qb = 6.68786$ | $n_\eta = 2$ |
|  | $m = \pm 2$ | $qR = 6.988$ | $n_\theta = 1$ | $qb = 6.79294$ | $n_\eta = 1$ |
|  | $m = \pm 3$ | $qR = 6.988$ | $n_\theta = 0$ | $qb = 6.92555$ | $n_\eta = 0$ |
| $l = 4$ | $m = 0$ | $qR = 8.183$ | $n_\theta = 4$ | $qb = 7.89418$ | $n_\eta = 4$ |
|  | $m = \pm 1$ | $qR = 8.183$ | $n_\theta = 3$ | $qb = 7.91407$ | $n_\eta = 3$ |
|  | $m = \pm 2$ | $qR = 8.183$ | $n_\theta = 2$ | $qb = 7.96667$ | $n_\eta = 2$ |
|  | $m = \pm 3$ | $qR = 8.183$ | $n_\theta = 1$ | $qb = 8.04388$ | $n_\eta = 1$ |
|  | $m = \pm 4$ | $qR = 8.183$ | $n_\theta = 0$ | $qb = 8.1373$ | $n_\eta = 0$ |

In Table 3 we compare the quantum states in a sphere and a spheroid with $a/b \approx 1$ for the case $n_r = 1$, taking into account that in a sphere the energy levels are degenerate in the quantum number $m$. The values of the reduced eigenvalues $qR$ for quantum states in a deep spherical well are taken from the Ref. 3. The values of the reduced eigenvalues $qb$ are determined for spheroids with aspect ratios $a/b \approx 1.06$. The indexes $n_\eta$ and $n_\theta$ denote the number of internal zeros for the angular function $S_{ml}(c,\eta)$ in the spheroid and for Legendre harmonics in a sphere, respectively. As Table 3 shows, in the limit $a/b \to 1$ for the same quantum numbers $m$ and $l$ the reduced eigenvalues $qb$ in the spheroid tend to the eigenvalues $qR$ in the sphere. However, the $qb$ values for the spheroid turn out to be slightly less than the $qR$ values, since in the calculations performed the aspect ratios of the spheroid are limited by the condition $a/b > 1.06$. Similar results were also obtained for quantum numbers $n_r = 2, 3$.

## 4. Conclusion

It should be noted that from the formal point of view, the theory of spheroidal functions is well developed and presented in various manuals, reference books and numerous original articles[6-19]. However, when performing specific calculations of eigenfunctions and eigenvalues of boundary value problem for the Helmholtz equation in a spheroid, a researcher or a student comes across a rather cumbersome mathematics. In this paper we show that the use of standard power series makes it easy to establish their convergence, to determine the eigenvalues of boundary value problems with high accuracy, and to study the behavior of angular and radial spheroidal functions in the relevant coordinate regions.

As an example of this approach, in this paper the low-lying energy levels of a deep spheroidal quantum well are calculated. The well-known results[1-5] for cylindrical and spherical wells are shown to be reproduced in the limits $a/b \gg 1$, and $a/b \to 1$, respectively.

## ACKNOWLEDGMENT

The author thanks Dr. Yu.V. Popov (Lomonosov Moscow State University) for stimulating discussions.

The author has no conflicts to disclose.